\documentclass[12pt]{iopart}
\usepackage{iopams}
\usepackage{setstack}
\usepackage{amsfonts}
\usepackage{amssymb}
\usepackage{dcolumn}
\usepackage{color}
\usepackage{bm}
\usepackage{graphicx}
\usepackage{epsfig}

\def \e{\mathrm{e}}
\def \i{\mathrm{i}}
\def \half{{1 \over 2}}
\def \da{^{\dagger}}
\def \la{\langle}
\def \ra{\rangle}
\def \g{\gamma}
\def \G{\Gamma}
\def \bG{\boldsymbol{\Gamma}}
\def \Gb{\bar{\Gamma}}
\def \bGb{\bar{\boldsymbol{\Gamma}}}
\def \pxipy{$p_x\!+\!ip_y$ }
\def \xvec{\boldsymbol{x}}
\def \kvec{\boldsymbol{k}}
\def \etavec{\boldsymbol{\eta}}
\def \hvec{\boldsymbol{h}_{\kvec}}

\def \dHk{\delta H_k}
\def \dEk{\delta E_k}
\def \pc{p_\mathrm{c}}
\def \pe{p_\mathrm{e}}
\def \Pf{\mathrm{Pf}}
\def \sign{\mathrm{sign}}

\begin{document}

\title{Majorana fermions on a disordered triangular lattice}
\author{
    Yaacov E.~Kraus and Ady Stern}
\address{
    Department of Condensed Matter Physics, Weizmann Institute of Science, Rehovot 76100, Israel}
\ead{kobi.kraus@weizmann.ac.il}

\begin{abstract}
Vortices of several condensed matter systems are predicted to have zero-energy core
excitations which are Majorana fermions. These exotic quasi-particles are neutral,
massless, and expected to have non-Abelian statistics. Furthermore, they make the ground state of the system highly degenerate. For a large density of vortices,
an Abrikosov lattice is formed, and tunneling of Majorana fermions between vortices removes the energy degeneracy. In particular the spectrum of Majorana fermions in a
triangular lattice is gapped, and the Hamiltonian which describes such a system is
antisymmetric under time-reversal. We consider Majorana fermions on a disordered triangular lattice. We find that even for very weak disorder
in the location of the vortices localized sub-gap modes appear. As the disorder
becomes strong, a percolation phase transition takes place, and the gap is fully
closed by extended states. The mechanism that underlies these phenomena is domain
walls between two time-reversed phases, which are created by flipping the sign of the
tunneling matrix elements. The density of states in the disordered lattice seems to diverge at zero energy.

\end{abstract}

\maketitle


\section{Introduction}
\label{Sec:Intro}

Exotic states of matter are among the most intriguing topics in the field of condensed matter physics.
One class of these exotic states are the two-dimensional (2D) systems in which
quasi-particles follow non-Abelian quantum statistics \cite{AdyNature}. The search
for such systems is driven both by their unique properties and by their potential
application to topological quantum computing \cite{RMP}.

In a non-Abelian state the ground state is degenerate when quasi-particles are
present, and the degeneracy increases exponentially with the number of
quasi-particles. Perhaps the simplest way of obtaining such a degeneracy is by
having quasi-particles that carry Majorana fermionic excitations \cite{ReadGreen}.
Majorana fermions (MFs) are expected to appear as zero-energy excitations in the
cores of vortices in a layered \pxipy superconductor -- such as proposed for
Sr$_2$RuO$_4$ \cite{Maeno}, in 2D systems that can be mapped onto such
a superconductor -- the $\nu = 5/2$ fractional quantum Hall state \cite{MooreRead},
on the surface of a topological insulator that is in proximity to an {\it s}-wave
superconductor and an insulating ferromagnet \cite{FuKane}, and at hybrid structures
of semiconductors and superconductors \cite{TewariSau}. Furthermore, they are
expected to form in 1D systems in proximity to $s$-wave superconductors
\cite{YuvalGil}, and in several other systems \cite{Sato,WimmerBeenakker}.

A variety of experiments have been proposed in order to probe the predicted MFs,
based on the their unique properties. To mention a few examples: zero-energy
excitations can be observed by performing STM measurements at the vortex core
\cite{PRL,PRB}; the degeneracy of the ground state affects the thermodynamical
properties \cite{thermodynamics1, thermodynamics2}; non-locality of an electron in a
Majorana state has a signature in tunneling between two vortices \cite{nonlocality1,
nonlocality2}; and interferometric experiments are able to probe the non-Abelian
statistics \cite{RMP, interferometry1, interferometry2, interferometry3}.

The suggested thermodynamics and interferometry measurements are based on the unique
many-body properties of the MFs, but assume that the MFs are localized at theirs
positions. The MFs appear at vortex cores, which are expected to rearrange as an
Abrikosov lattice at high enough density. The lattice order and the small tunneling
amplitude between neighboring vortices remove the ground state degeneracy and form a
band of low-energy excitations. On one hand, this band may serve to  probe the
existence of the MFs. On the other hand, it may conceal signals of suggested
many-body measurements, especially controlled adiabatic processes, such as
interferometry.

Some previous works have analyzed clean periodic square, triangular \cite{EytanAdy} and
honeycomb \cite{KitaevModel} lattices of MFs, and found the electronic  conductivity
associated with tunneling between vortex cores. Other works have considered some
aspects of random square \cite{WimmerBeenakker} and honeycomb \cite{Lahtinen}
lattices. In this paper, we consider disordered triangular lattices. This lattice
breaks time-reversal symmetry: it may be mapped onto a tight-binding model of
electrons on the same lattice with each plaquette being pierced by $\pm 1/4$
magnetic flux quantum. The sign of the flux is determined by the sign of the tight-binding coupling term. The spectrum of the perfect lattice is gapped.

Our main finding is that disorder in the \emph{sign} of the tunneling amplitude
between neighboring sites creates a peak in the density of states (DOS) at zero
energy. We show that at weak disorder zero-energy localized sub-gap states are created, and coupling between these states creates a DOS close to zero energy. At strong disorder the sign flipping creates domain walls between regions of the two
time-reversed phases, and chiral modes appear along these walls. When the signs are random, we find the domains to be narrow and the spectrum of excitations to be characterized by strong dependence
on the geometry of the walls. Plausibly, that is the source of the zero-energy DOS
peak in the highly disordered system. The
transition between these two limits is a percolation phase transition, taking place when the probability of flipping a sign is around 0.15. We find the DOS to diverge at zero energy at intermediate and strong disorder.

The paper is organized as follows: In section \ref{Sec:Disorder}, we define the MFs
lattice model, and show the numerical DOS for a disordered lattice. In section
\ref{Sec:Edge}, we examine how low-energy excitations may emerge from localized modes
in the limit of weak disorder and from extended interfaces in the limit of large
$p$. Section \ref{Sec:Percolation} discusses the dependence of the excitation
spectrum on the strength of the disorder. Section \ref{Sec:Summary}
summarizes the results and compares them to previous works.


\section{A disordered triangular lattice}
\label{Sec:Disorder}

Our interest here is in MFs that form a triangular lattice. An MF is defined as a self-adjoint operator $\g = \g^\dag$ that satisfies
fermionic anti-commutation relation $\{ \g_i, \g_j \} = \delta_{ij}$. A standard
complex fermionic operator can be constructed from MFs by superposing an even number of them. For
example $\psi_n = (\g_i + \i \g_j) / \sqrt{2}$ for two MFs satisfies the standard
anti-commutation relations $\{ \psi_n, \psi_m^\dag \} = \delta_{nm}$ and $\{ \psi_n,
\psi_m \} = 0$.

We consider MFs that are solutions of the Bogoliubov de-Gennes (BdG) equation in the presence of vortices in the superconductor. Each vortex in the superconductor carries a signle MF, whose wavefunction
is exponentially localized around the vortex's core. Overlaps of the MF wavefunctions of neighboring vortices result in tunneling matrix elements. Vortices in superconductors and quasi-particles in clean quantum Hall systems are
arranged on a lattice; thus the MFs are also arranged as a lattice, and their dynamics may naturally be described by the tight-binding model.

The particle-hole symmetry of the BdG equation implies that a single MF is a zero-energy solution. Hence, in the tight-binding Hamiltonian the on-site energy of the MFs is zero, and there are only hopping terms $t_{ij}$. Moreover, for the Hamiltonian to be Hermitian, $t_{ij} \g_i \g_j = (t_{ij} \g_i \g_j)^\dag = -t_{ij}^* \g_i \g_j$, implying a purely imaginary hopping term. By assuming discrete symmetry of the lattice to translations, we can write the simple Hamiltonian
\begin{equation}  \label{Eq:tightbinding}
    H = \i t \sum_{\la ij \ra} s_{ij} \g_i \g_j,
\end{equation}
where $\la ij \ra$ are nearest neighbors and $s_{ij} = -s_{ji} = \pm 1$.

Any element $s_{ij}$ is gauge dependent, because the $\g_i$ operators are defined up
to an overall sign. However, the product of $s_{ij}$'s along a path creating a
closed loop is gauge independent. It has been shown in \cite{EytanAdy} that for any
lattice whose plaquette is a polygon of $n$ vertices, the product of $s_{ij}$ around each plaquette corresponds to the plaquette enclosing $n/4 -
1/2$ flux quanta. Therefore the product of the hopping terms along the bonds that
create the plaquette is $-\i^n t^n$, and the product of the $s_{ij}$'s along this
path is $-1$ (the direction of the path is chosen to be aligned with the chirality
of the order parameter).

\begin{figure}[htb]
\begin{center}
\vspace{0cm}
\includegraphics[width=12cm,angle=0]{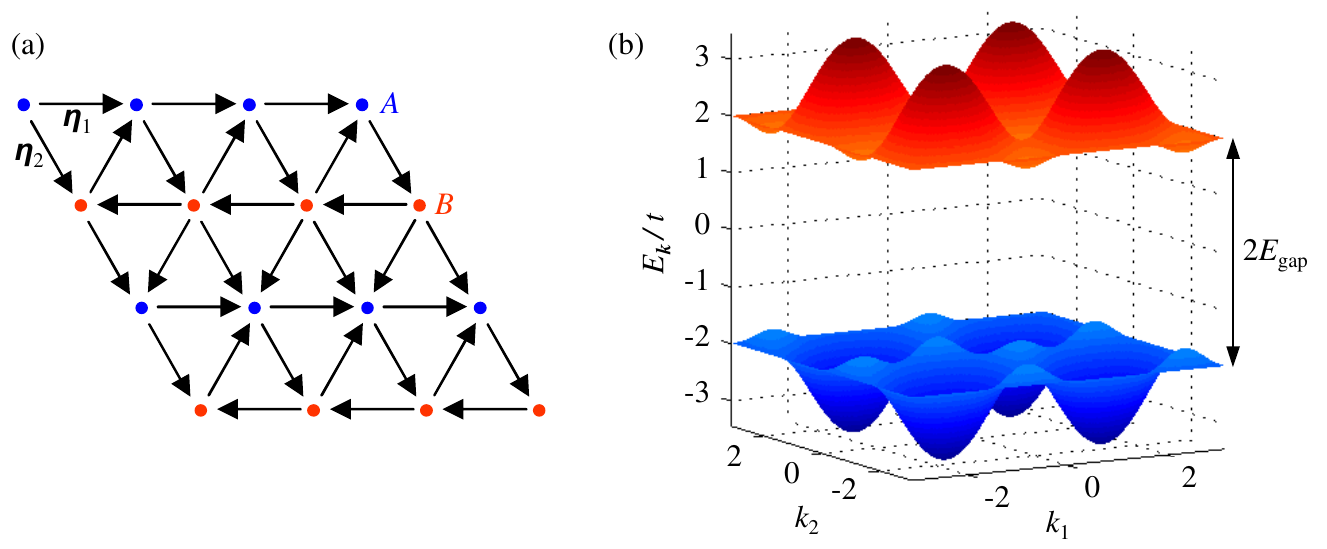}
\vspace{0cm} \caption{ \label{Fig:triangular} %
(a) One possible gauge of the tight-binding model of MFs on a triangular lattice, which is expressed in equation
(\ref{Eq:Hclean}). An arrow from site $i$ to site $j$ means $s_{ij} = 1$. Note that
the lattice is split into two sublattices $A$ and $B$, and that the counterclockwise
product of the $s_{ij}$'s around each triangle equals to $-1$. (b) The corresponding
gapped spectrum, with $E_\mathrm{gap} = 1.73$.}
\end{center}
\end{figure}

In particular, in the triangular lattice each plaquette encloses a quarter of
flux quantum, and the product of the hopping terms equals $\i t^3$, revealing a time-reversal anti-symmetry. Moreover, the unit cell which encloses a flux quantum is a parallelogram of four neighboring triangles, leading to the lattice being
composed of two sublattices. One possible gauge is illustrated in figure
\ref{Fig:triangular}(a), which also depicts the two sublattices $A$ and $B$. In this
gauge
\begin{eqnarray}  \label{Eq:Hclean}
    H & = \i t \sum_{i} \left( -\g_{A,i} \g_{A,i + \etavec_1} + \g_{B,i} \g_{B,i + \etavec_1}
                          + \g_{B,i} \g_{A,i - \etavec_2} \right. \nonumber \\
      & \qquad\qquad \left. + \g_{A,i} \g_{B,i - \etavec_2} - \g_{B,i} \g_{A,i + \etavec_1 - \etavec_2}
                          + \g_{A,i} \g_{B,i + \etavec_1 - \etavec_2} \right).
\end{eqnarray}
Let us assume a clean periodic lattice of $L_1 \times L_2$ sites, with $L_2$ even. We can define
the Fourier transform $\Gb_{a,\kvec} = \sum_i \e^{-\i \kvec \xvec_i} \g_{a,i}$,
where $a = A,B$ and $\kvec = 2\pi( m_1/L_1 , m_2/2L_2 )$ for $m_i = 0, ..., L_i-1$
($i = 1,2$) \cite{EytanAdy}. These transformed operators are complex fermions, which
satisfy $\{ \Gb_{a,\kvec}, \Gb_{a,\kvec'}^\dag \} = \delta_{\kvec \kvec'}$. Note,
however, that for $\kvec = (0,0)$, $(\pi,0)$, $(0,\pi/2)$ and $(\pi,\pi/2)$ they are
MFs. By denoting spinor $\bGb_{\kvec}\da = ( \Gb_{A,\kvec}\da, \Gb_{B,\kvec}\da )$
the Hamiltonian can be expressed as
\begin{equation}  \label{Eq:Hk1k2}
H = 2t \sum_{\kvec} \bGb_{\kvec}\da \left( \sin k_2 \sigma_x - \cos(k_1 - k_2)
\sigma_y - \sin k_1 \sigma_z \right) \bGb_{\kvec},
\end{equation}
where the Pauli matrices act on the sublattice space. The resulting spectrum has an
energy gap of $2tE_\mathrm{gap}$, where $E_\mathrm{gap} = 1.73$, as depicted in
figure \ref{Fig:triangular}(b).

The hopping elements between neighboring MFs are very sensitive to the
inter-vortex separation. For example it was shown that for a \pxipy
superconductor \cite{Cheng}:
\begin{equation}  \label{Eq:t_vs_r} %
    t\sim \cos\left( k_\mathrm{F} r + {\pi \over 4} \right) \e^{-r/\xi},
\end{equation}
where $k_\mathrm{F}$ is the Fermi momentum, $r$ is the inter-vortex distance, and $\xi$ is the coherence length, which
is usually larger than $k_\mathrm{F}$. The exponential decay and the oscillations were found to occur also for the $\nu=5/2$ case \cite{Baraban}, and are likely to appear in all realizations. Therefore small deformations of the
Abrikosov lattice of order $k_\mathrm{F}^{\phantom{f}-1}$ will produce fluctuations
in both the amplitude and sign of the hopping terms $t_{ij}$.

\begin{figure}[htb]
\begin{center}
\vspace{0cm}
\includegraphics[width=12cm,angle=0]{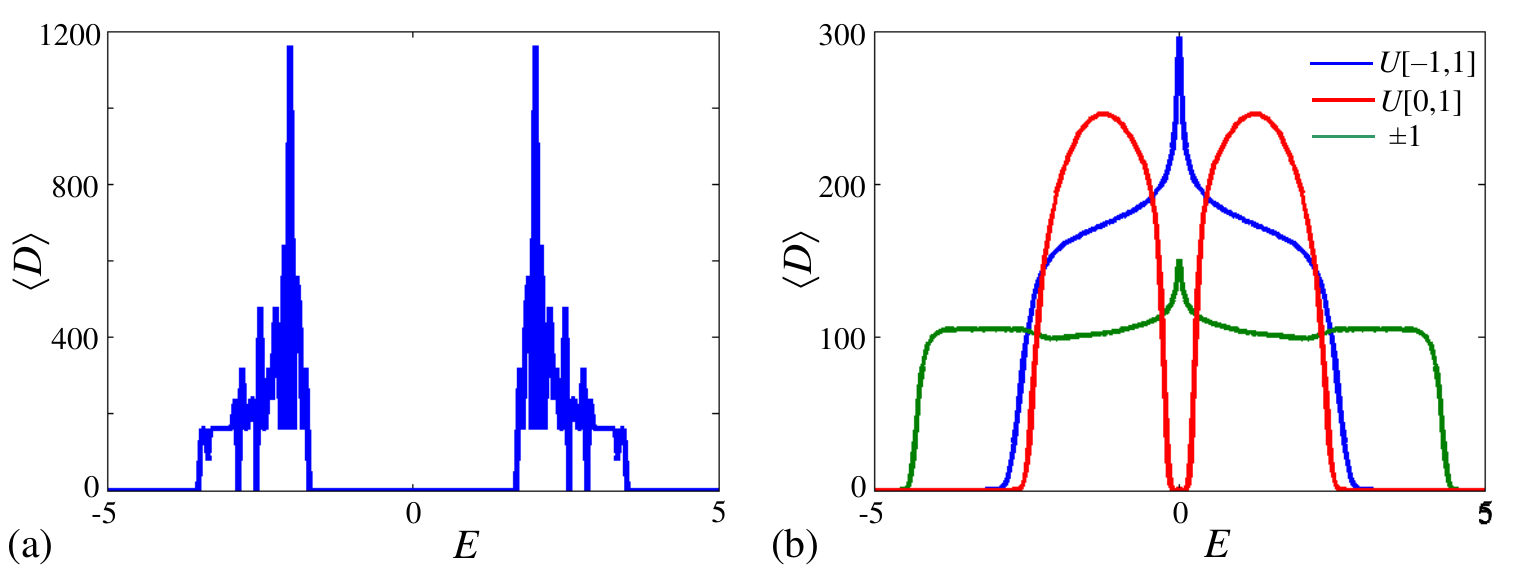}
\vspace{0cm} \caption{ \label{Fig:DOSall} %
The disorder-averaged DOS $\la D \ra$ as a function of the energy $E$ of a periodic $30 \times
30$ lattice with (a) uniform and (b) random hopping terms. In systems with hopping terms that are random both in sign and in magnitude
(blue) and in systems where only the signs are random (green) there is a sharp
peak at zero energy, while random amplitudes (red) only smear the spectrum of the
clean system, but do not close the gap. The DOS is averaged over 10,000
realizations.}
\end{center}
\end{figure}

We find numerically that random hopping terms $t_{ij} = ts_{ij}$, where $s_{ij}$ is uniformly distributed in
the interval $[-1, 1]$, close the energy gap of the uniform system. The DOS, which
appears in figure \ref{Fig:DOSall}, shows that not only the gap is closed, but a
peak emerges at zero energy. We can distinguish between random amplitudes and random
signs of the hopping terms, i.e. $|s_{ij}| \sim U[0, 1]$ or $s_{ij} = \pm1$ in equal
probability, respectively. The DOS of these two cases, which are also depicted in
figure \ref{Fig:DOSall}, clearly shows that the zero-energy peak in the DOS appears
only due to random signs, while random amplitudes merely smear the spectrum without
closing the energy gap.
We are mostly interested in the zero-energy peak of the DOS, and therefore in the
following we will focus on the case where disorder appears in the sign of the hopping terms.

\begin{figure}[htb]
\begin{center}
\vspace{0cm}
\includegraphics[width=13cm,angle=0]{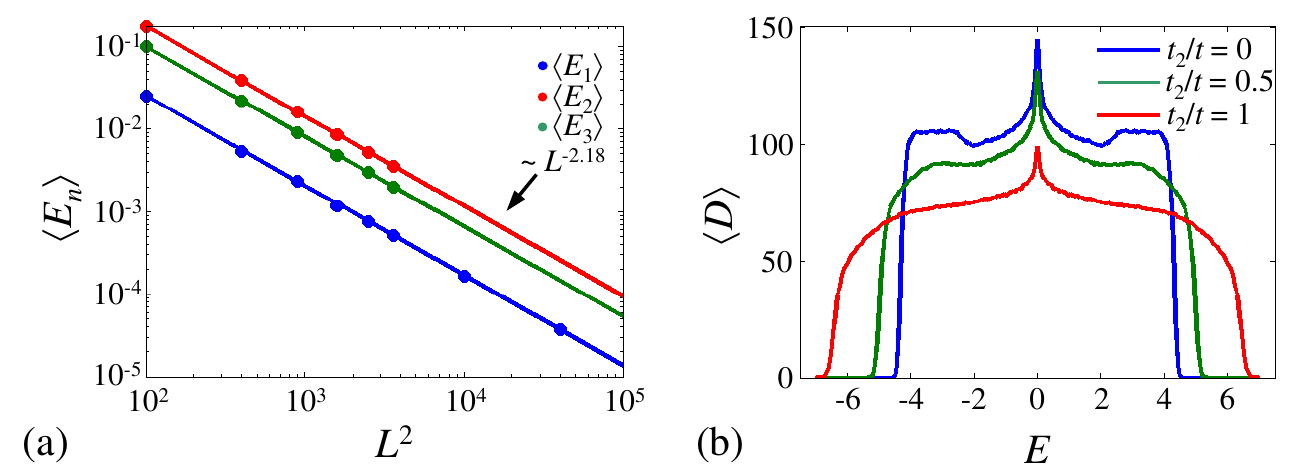}
\vspace{0cm} \caption{ \label{Fig:EnL} %
(a) The dependence of the disorder-averaged energies of the three lowest states $E_1, E_2$ and $E_3$,
as a function of the system size $L^2$, on a log-log scale. The energies $\la E_n \ra$
were averaged over 1000 realizations. Linear fitting gives $\la E_n \ra \propto
L^{-2.18}$ for all three energies. (b) The disorder-averaged DOS $\la D(E) \ra$ with first
($t$) and second ($t_2$) nearest neighbors hopping, both with random signs, for
$t_2/t = 0, 0.5, 1$. As $t_2$ approaches $t$ the DOS approaches a semicircle. The DOS
is of a $30 \times 30$ lattice, and was averaged over 1000 realizations. }
\end{center}
\end{figure}

The sharp peak that we found numerically for the density of states of finite systems close to zero-energy naturally raises the question of the way the DOS scales with the size of the system in the thermodynamic limit. Increasing the system size $L$ obviously increases the DOS. If the lowest
energies decay faster than $1/L^2$, then the zero-energy DOS will increase faster than
$L^2$, and the zero-energy DOS per unit area would diverge in the thermodynamic limit. Figure \ref{Fig:EnL}(a) exhibits the energy of the three lowest states $E_1,
E_2$ and $E_3$, averaged over disorder, as a function of the system size. In the given range a fit to $\la E_n \ra
\propto 1/L^{\alpha}$ yields $\alpha\approx 2.18$ for $n = 1,2$ and $3$, which indicates either a weak power-law or logarithmic divergence of the DOS per unit area.

It is instructive to compare the DOS we find for the case of random nearest-neighbor
hopping to that we find for the case of a random Hermitian matrix of imaginary
terms. The DOS of the latter is a version of a semicircle distribution,
with a Delta peak at zero energy, due to the symmetry of the spectrum with respect
to reflection about zero energy \cite{KalischBraak}. When we add second nearest-neighbor hopping with random signs to the Hamiltonian, the DOS indeed gets closer to the semicircle. Figure \ref{Fig:EnL}(b) shows the spectrum for several ratios of
$t_2$, the amplitude of the next-nearest-neighbor terms, to $t$. We note, however,
that we numerically find the gap to close even for the clean system when $t_2/t
\approx 0.58$.

\begin{figure}[htb]
\begin{center}
\vspace{0cm}
\includegraphics[width=6.5cm,angle=0]{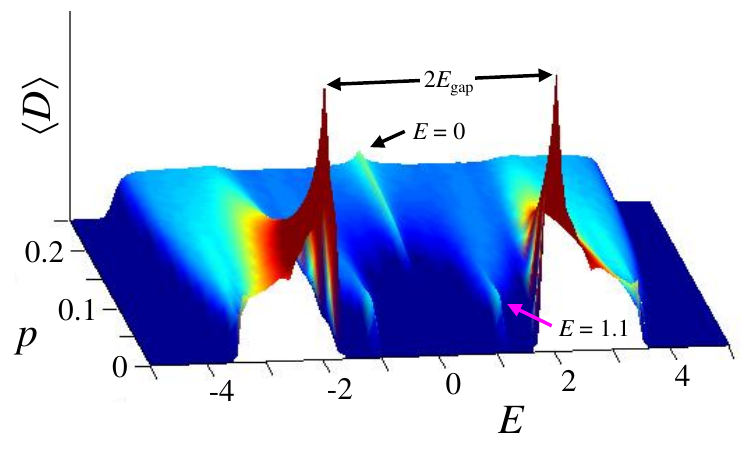}
\vspace{0cm} \caption{ \label{Fig:DOS_vs_p_t2} %
The disorder-averaged DOS $\la D(E) \ra$ of a periodic $70 \times 70$ lattice as a function
of the probability $p$ for flipping the sign of a hopping term. For $p = 0$ the DOS is
that of figure \ref{Fig:DOSall}(a). Two small peaks at $E \approx \pm1$, that grow
linearly with $p$, belong to localized states which are created around a single
isolated flipped hopping term. The zero-energy peak grows slowly, but survive at the disordered
system. The DOS is averaged over 100 realizations. }
\end{center}
\end{figure}

Having established that the DOS of the clean lattice is characterized by an energy gap limited
by two singular peaks, and the lattice where the sign of the hopping terms is random is characterized by a zero
energy peak above a gapless background, we examine the evolution of the DOS with $p$,  the probability of
flipping the sign of a hopping term, which increases from 0 (the clean lattice) to 0.5 (the random sign limit). Figure \ref{Fig:DOS_vs_p_t2}(a) shows how increasing $p$  makes the singular peaks of the
clean system spread, and gradually creates sub-gap states.

In the next section we study the way the gap is closed, and especially the
appearance of the zero-energy peak. We address these questions by an examination of two limits. In the weak disorder limit, we present the minimal disorder configuration that results in a zero-energy state. In the strong disorder limit, we examine the low-energy states that are formed along domain walls between regions in which the signs of flux piercing the plaquettes are opposite.


\section{The origin of the zero-energy peak in the density of states}
\label{Sec:Edge}

We have seen that the clean triangular lattice of MFs is anti-symmetric under time
reversal, and has an energy gap. Depending on the sign of the hopping tunneling elements, the MFs may be mapped onto electrons in a triangular lattice pierced by $1/4$ or $-1/4$ flux quanta per plaquette. These two cases are characterized by two opposite Chern numbers:  by writing the $k^{th}$ component of equation (\ref{Eq:Hk1k2}) as $H_k = 2t \hvec \cdot \boldsymbol{\sigma}$, it is easy to see that the Majorana band is characterized by a non-trivial Chern number
\begin{eqnarray} \label{Eq:Chern}
\nu & = & \sign(t) \int \frac{\mathrm{d}^2 k}{4\pi} \frac{1}{|\hvec|^3} \left( \hvec
\cdot \frac{\partial \hvec}{\partial k_1} \times \frac{\partial \hvec}{\partial k_2}
\right) \nonumber \\
& = & \sign(t) \int_{-\pi}^{\pi} \mathrm{d}k_1 \int_{-\pi/2}^{\pi/2} \mathrm{d}k_2
\frac{\sin^2(k_1 - k_2) + \cos k_1 \cos k_2 \cos(k_1 - k_2)}{4\pi \left[ \sin^2 k_1
+ \sin^2 k_2 + \cos^2(k_1 - k_2) \right]^{3/2}} \nonumber \\
& = & \sign(t).
\end{eqnarray}
The disorder-induced reversal of signs of hopping terms may revert the sign of the flux in some of the plaquettes in the lattice. The systems would then have islands of one phase separated by lines of interface from the bulk of the other phase.
An infinite interface between two phases of different Chern numbers is accompanied by gapless modes \cite{Hatsugai,Wen,Gils}. For finite islands, one may naively expect finite-size quantization to induce a gap in the energy spectrum, and thus low-energy excitations to require large islands. This expectation is only partially valid. As we explain below, for the Majorana lattice, large islands are associated with low-energy excitations at their edges, but small islands may carry low-energy and zero-energy modes as well.

When the probability $p$ of flipping the sign of a hopping term is very small the flipped terms are dilute, and most of them are isolated. A two-triangles island, which is created by flipping the sign of a single hopping term, results in two localized sub-gap states with energies of approximately $\pm 1.1$. Theses states are the
source of the peaks in the DOS at $\pm E \approx \pm 1.1$, and indeed in the limit of small probability the DOS associated with these peaks  $\la D(E \approx \pm 1.1) \ra \propto p$. Such islands do not, however, contribute to the DOS close to zero energy.

As we show in \ref{App:Mininal}, flipping three hopping terms with a common vertex creates two localized states with energy that is either zero or exponentially small with the system size. We also show that this is the minimal way of creating zero modes. Thus, for small $p$ the zero-energy DOS is dominated by the probability of creating such configurations, and should scale like $p^3$. The two modes are, however, split when there is a single flipped hopping term at a distance $r$ from that vertex.  The exact splitting depends on the orientations of the bonds with flipped signs, but we found numerically that it can be well approximated by $\pm E_\mathrm{t}(r) \approx \pm 6 \e^{-2r}$. For $E_\mathrm{t}$ to be much smaller than the typical finite-size energy splitting, that scales as $L^{-2}$, we need $r\gg \log{L}$. A given radius $r$ encloses approximately $3\pi r^2$ hopping terms around the vertex, and the probability that all these terms are unflipped is $(1-p)^{3\pi r^2}$. Thus, the $p^3$ dependence of the zero-energy DOS is limited to small values of $p\ll \log^{-2}L$. For example, for a $70 \times 70$ lattice we get $p < 0.01$.

As $p$ gets large, the plaquettes in which the flux is reversed connect to one another, and the system is split into domains with opposite fluxes. To understand the excitations spectrum in this limit, we first examine the spectrum of excitations associated with the various possible interfaces of two large domains of opposite fluxes. Then, we examine the statistical distribution of such interfaces as a function of the flipping probability $p$.

At the interface between a macroscopic region of Chern number $\nu = \pm 1$ and the vacuum, a chiral gapless mode must appear \cite{Hatsugai,Wen,Gils}. In order to explicitly find this mode, we first replace $\etavec_1$ and
$\etavec_2$ in equation (\ref{Eq:Hclean}) by $\hat{x}$ and $\hat{y}$ for simplicity,
and denote the lattice sites by $\boldsymbol{x}$. We assume rotational symmetry
along $\hat{y}$, and define the operators $\G_{a,kx} = \sum_y \e^{-\i ky}
\g_{a,\xvec}$ with $a = A,B$ and $k = \pi m/L_y$, where $m = 0, ..., L_y-1$. Each
$\G_{a,kx}$ represents a wavefunction that is localized at $\hat{x}$ but extended at
$\hat{y}$. With these operators the Hamiltonian becomes
\begin{equation}  \label{Eq:Hk}
\fl H = -t \sum_{k,x} \left[ \bG_{k,x}\da (\e^{-\i k}\sigma_y + \i \sigma_z)
\bG_{k,x+1}
 + \bG_{k,x}\da (\e^{\i k}\sigma_y - \i \sigma_z) \bG_{k,x-1}
     - \bG_{k,x}\da (2\sin k\sigma_x) \bG_{k,x} \right].
\end{equation}

Given a right edge at $x = 0$, the edge eigenmode of (\ref{Eq:Hk}) for $x \leq 0$ is
$\sum_{n=0}^{\infty} b_k^{\phantom{k}-2n} (1,1) \bG_{k,-2n}$, where $b_k = \i
\e^{-\i k} \tan (k/2) \approx \i \half k \ll 1$ and $(a,b) \bG_{k,x} = a\G_{A,k,x} + b\G_{B,k,x}$. For $k=0$ it reduces to  $(1,1) \bG_{0,0}$. The dispersion of the edge mode is
$E_k = 2\sin k$, with $E_k \approx 2k = 2\pi m/L_y$ at low energy. Note that the
wavefunction is localized in the $\hat{x}$ direction with a localization length of
the order of $1/\log(L_y/\pi m)\sim 1/\log{ E}$, which depends weakly on the energy and the system
size.

\begin{figure}[htb]
\begin{center}
\includegraphics[width=8cm,angle=0]{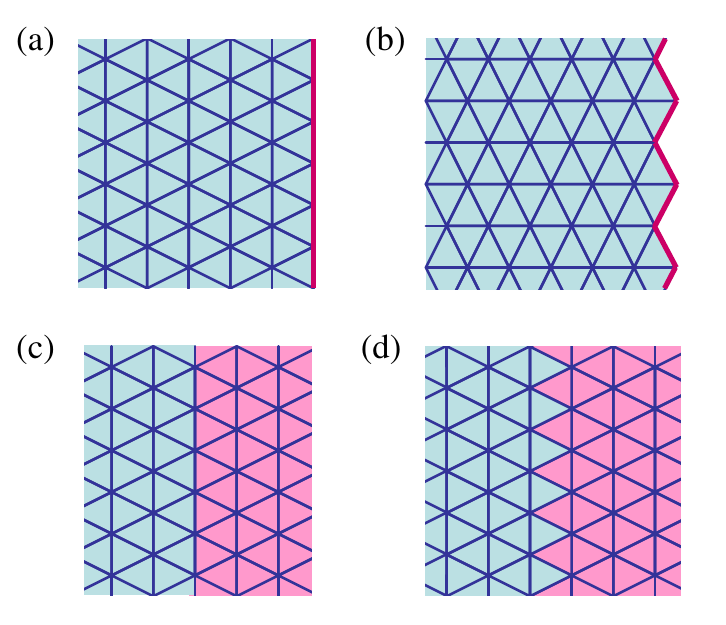}
\caption{ \label{Fig:edge_types} %
Two types of edges in the triangular lattice with one periodic coordinate: (a) flat
and (b) zigzag. With the periodic coordinate of (a) there are two types of domain
walls between P-phase (azure) and N-phase (pink) that preserve the translational
symmetry: (c) flat and (d) sawtooth. }
\end{center}
\end{figure}

A triangular lattice with one periodic coordinate can have either flat or zigzag
edges, depending on the way the periodic boundary condition is taken, as illustrated
in figure \ref{Fig:edge_types}. The Hamiltonian (\ref{Eq:Hk}) describes a flat edge. In
a zigzag edge, we also get a chiral edge mode, with the dispersion $E_k \approx
\sqrt{2}k$ and $b_k \approx ( 3 - 2\sqrt{2} ) - ( 8 - 11/\sqrt{2} ) k^2$, which
gives a localization length that is approximately independent of energy, at low energies. We can see that in spite of the
differences, both edges show a similar behavior of linear dispersion and
exponentially localized wavefunction. Henceforth we will assume the periodicity of
equation (\ref{Eq:Hk}).

An interface, or domain wall, between the two phases of opposite Chern numbers is expected to hold
two chiral edge modes, similar to the single mode that exists along the interface
between any of the two phases and the vacuum.
We denote the unflipped phase by P and the flipped phase by N, representing
their positive and negative fluxes respectively. A periodic lattice with an interface between a P-phase and an N-phase may have two kinds of domain walls that preserve the translational symmetry: flat and sawtooth, as depicted in figure
\ref{Fig:edge_types}.

A flat domain wall is created when $\sigma_y \rightarrow -\sigma_y$
in the hopping terms for $x > 0$. The two chiral modes of such a domain wall have
the same dispersion and amplitudes $b_k$ as the flat edge mode, where in the N-phase
the amplitudes are $b_k^{\phantom{k}*}$, as expected from the time reversal. Only
now one mode has non-vanishing amplitudes at $x=2n$, while the other mode has
non-vanishing amplitudes at $x=2n+1$. A sawtooth domain wall is a result of flipping
the $\sigma_x$'s. It shows essentially a similar behavior to the flat domain wall,
but with $E_k \approx 1.95( k \pm k_\mathrm{z} )$, where $k_\mathrm{z} = \arccos(
\sqrt{5}/2 - 1/2 )$. Again, we see that the geometry of the domain walls has only a small
effect on the low-energy physics.

If an `island' of a roughly circular shape of the N-phase is surrounded by the P-phase
lattice, then the two chiral modes will be localized along the perimeter of the
island, and their dispersion will be $E_m \sim m/M$, where $M$ is the perimeter of
the island. Such an island will close the energy gap only if it becomes macroscopic.

\begin{figure}[htb]
\begin{center}
\includegraphics[width=14cm,angle=0]{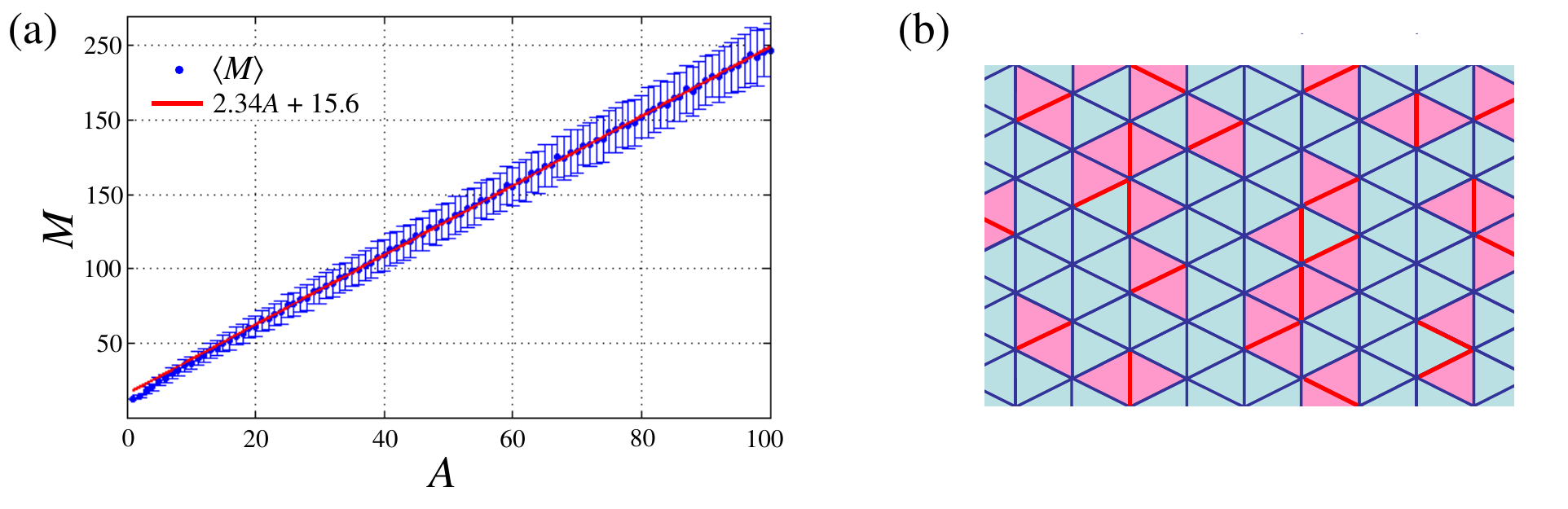}
\caption{ \label{Fig:perimeter_area} %
(a) The distribution of the perimeters $M$ of the N-phase islands with a given area
$A$, where the islands were created by hopping terms with random signs. The error
bars are located at the mean values $\la M \ra$, with the width of a standard
deviation. The mean value is excellently fitted to a linear curve (red line). The
distribution was taken from 10,000 random realizations of a $30 \times 30$ lattice.
(b) A typical islands configuration of N-triangles (pink) that are created by
random flipping of hopping terms (red), with a complicated geometry of rhombi and
triangles. In this realization 0.14 of the hopping terms are flipped. }
\end{center}
\end{figure}

In a disordered system where the sign of the hopping term $s_{ij}$ is random, islands of the N-phase will be created in various shapes and sizes. A way to parameterize the morphology
of an island is by the perimeter-area relation. In approximately rounded shapes $M
\propto A^{1/2}$, where $A$ is the area of the island, while in a narrow strip $M
\propto A$. Figure \ref{Fig:perimeter_area}(a) depicts the distribution of the
perimeters as a function of the areas of random islands in a $30 \times 30$ lattice, where the probability of flipping a sign is $1/2$.
The area of an island is the number of N-triangles that compose the island,
where two N-triangles are defined to belong to the same island if they share at
least a vertex. The perimeter is composed of the P-triangles which have a common
vertex with the island. For islands with $A > 10$ the mean perimeter can be
excellently approximated by $\la M(A) \ra \approx 2.34 A + 15.6$, which means that
the islands look like snakes or gossamer.

The smallest island that may be created by flipping the sign of a single hopping term is composed of two triangles, and has a rhombus shape. The simplicity of the shapes is not preserved for larger islands. Flipping two neighboring hopping terms
results in two flipped N-triangles, but those share only a vertex. Larger islands are typically of complicated shapes, composed of rhombi and triangles, as illustrated in figure \ref{Fig:perimeter_area}(b). Complicated islands provide
complicated dispersion and DOS, as we will demonstrate. In order to get a feeling to the resulting dispersions, we now focus on three types of narrow islands that preserve the translational symmetry, i.e.~strips of N-triangles.

The method we use for approximating the low-energy dispersion of the strips is first
to find zero-energy modes with $k = 0$, and then extract the dominant $k$-dependence
employing first order perturbation theory. For the clean system the Hamiltonian (\ref{Eq:Hk}) may be recast as \numparts
\begin{eqnarray}
\fl H & = & H_0 + \sum_{k \neq 0} \dHk, \\
\fl H_0 & = & 2t \sum_{x} \bG_{0,x} (-\sigma_y + \i \sigma_z) \bG_{0,x+1},  \label{Eq:H0} \\
\fl \dHk & = & 2t\sin{k \over 2} \sum_{x} \left( \e^{-\i k/2} \bG_{k,x}\da \i
\sigma_y \bG_{k,x+1}  - \e^{\i k/2} \bG_{k,x}\da  \i \sigma_y \bG_{k,x-1}
     + 2\cos{k \over 2} \bG_{k,x}\da \sigma_x \bG_{k,x} \right).  \label{Eq:dHk}
\end{eqnarray}
\endnumparts
For any strip $\mathrm s$, this Hamiltonian has to be modified to the Hamiltonian $H^\mathrm{s}$, which may be decomposed similarly to $H_0^\mathrm{s}$ and $\delta H_k^\mathrm{s}$. The $k=0$ zero modes of $H^\mathrm{s}$ will be denoted by $\G^\mathrm{s}_i = \G^{\mathrm{s}\dag}_i$. They satisfy $[H_0^s, \G^\mathrm{s}_i] = 0$. In the cases we will explicitly consider here, there are two zero modes per strip.  The matrix elements of $\dHk$ between these zero modes give the leading order of the $k$-dependence. Therefore after finding $\G^\mathrm{s}_\mathrm{I}$ and $\G^\mathrm{s}_\mathrm{II}$, we diagonalize the matrix $[\dHk]$, where $[\dHk]_{ij} = \{\G^\mathrm{z}_i, [\dHk, \G^\mathrm{z}_j]\}$

\begin{figure}[htb]
\begin{center}
\includegraphics[width=9cm,angle=0]{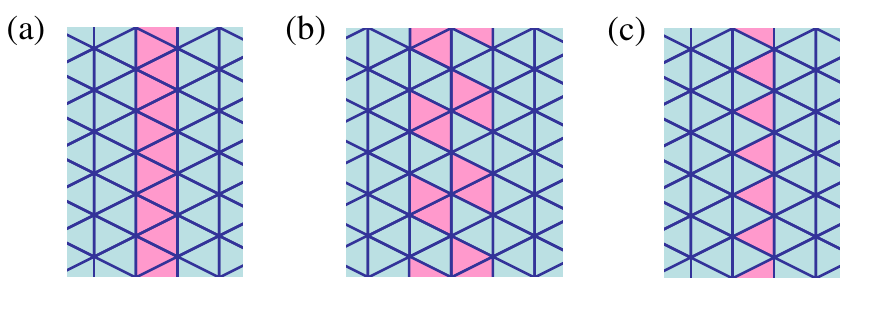}
\caption{ \label{Fig:strips} %
The three examples of narrow islands of N-phase, whose chiral sub-gap modes
show three kinds of dispersion: (a) a flat strip with $E_k \propto k$, (b) a `zipper'
strip with $E_k \propto k^2$ and (c) a sawtooth strip with $E_k \propto k^3$. }
\end{center}
\end{figure}

The first and simplest strip is the flat strip, which is depicted in figure
\ref{Fig:strips}(a), and is created by flipping the sign of the $\sigma_z$'s in equation (\ref{Eq:H0}) between
$x = 0$ and $x = 1$. Now $H_0^{\mathrm{flat}}$ has two zero modes:
\begin{eqnarray} \label{Eq:bGz_flat}
\G_\mathrm{I}^\mathrm{flat}  & = & {1 \over \sqrt{2}} (1,1) \bG_{0,0}, \nonumber \\
\G_\mathrm{II}^\mathrm{flat} & = & {1 \over \sqrt{2}} (1,-1) \bG_{0,1}.
\end{eqnarray}
The matrix elements of $\dHk$ in the zero modes space are
\begin{equation} \label{Eq:dHk_flat}
[\dHk]^\mathrm{flat} = -t\sin{k}  \left( 1,-\tan{k \over 2},2 \right) \cdot
\boldsymbol{\tau},
\end{equation}
with $\boldsymbol{\tau}$ being the vector of Pauli matrices. The resulting
corrections to the zero energies are $\dEk^\mathrm{flat} \approx \pm
\sqrt{5}tk$. These modes are essentially the domain wall modes, which are modified
due to their hybridization, and similarly their DOS is a constant.

The second example is a `zipper' strip, shown in figure \ref{Fig:strips}(b), which
is a result of replacing $\sigma_y$ of Equations (\ref{Eq:H0}) and (\ref{Eq:dHk}) by $\i\sigma_x$ for $x=-1$ and by $-\i\sigma_x$ for $x=0$. Two zero modes again appear
\begin{eqnarray} \label{Eq:bGz_zip}
\G_\mathrm{I}^\mathrm{zip} & = & \frac{1}{2\sqrt{2}} \left[ (1,1) \bG_{0,-2} +
(0,2) \bG_{0,0} - (1,-1) \bG_{0,2} \right], \nonumber \\
\G_\mathrm{II}^\mathrm{zip} & = & \half \left[ (1,1) \bG_{0,-1} + (1,-1) \bG_{0,1}
\right],
\end{eqnarray}
which yields $[\dHk]^\mathrm{zip} = -\sqrt{2}t\sin^2{k \over 2} \sigma_y$. The
corrections to the energies are now quadratic $\dEk^\mathrm{zip} \approx \pm {1
\over \sqrt{8}}tk^2$. Note that these modes are wider, and that their DOS diverge as
$E^{-1/2}$.

The third example is the sawtooth strip, which is created by multiplying the hopping
terms between $x=-1$ and $x=0$ by $\sigma_z$, and is shown in figure \ref{Fig:strips}(c). The zero modes are now
\begin{eqnarray} \label{Eq:bGz_sawtooth}
& \G_\mathrm{I}^\mathrm{saw}  = {1 \over \sqrt{2}} (1,-1) \bG_{0,0}, \nonumber \\
& \G_\mathrm{II}^\mathrm{saw} = \half \left[ (1,1) \bG_{0,-1} + (1,-1) \bG_{0,1}
\right].
\end{eqnarray}
Unfortunately, $[\dHk]^\mathrm{saw} = -t(1+\tau_z)\sin{k} $, which means that the
first order perturbation theory gives the correction only to the first mode
$\dEk^\mathrm{saw} \approx -2tk$, while the second mode requires higher orders. An
explicit solution of this mode gives $\dEk^\mathrm{saw} \approx k^3$; thus this
soft mode makes the DOS diverge as $E^{-2/3}$.

We can conclude from this section that along edges and domain walls chiral modes
appear with linear low-energy dispersion. These modes close the energy gap only for
macroscopic domain walls, which will give a constant DOS. However, when the domains
become narrow strips, the dispersions of the chiral modes strongly depend on the
geometry of the strip, and their DOS tends to diverge at zero energy. And since most
domains which are created by random hopping terms are composed of narrow strips,
this may explain the zero-energy peak raising above the uniform background in the
DOS.


\section{Percolation phase transition}
\label{Sec:Percolation}

In the previous section we have seen that chiral modes
which are localized along stripe-shaped domains may lead to a zero-energy peak
in the DOS when the signs of the  hopping terms are random. Such modes provide low-energy states (states whose energy scales inversely with the size of the system) only if the length of the domain is of the order of the system size.
For hopping terms whose signs are random ($p=0.5$) there are such extended domains, since triangles of
both phases are distributed all over the lattice. In contrast, for small $p$ only small isolated domains are created.

In this section we address the dependence of the DOS on $p$. We start by examining the dependence of the size of the islands on $p$. It is natural to expect a percolation phase transition, where below a critical probability $\pc$ only microscopic islands can be found in the lattice. The probability distribution to find an island
with area $A$ is then exponential with $A$, where the typical area depends on $p$ and is independent of the
system size $L$. Above $\pc$ a macroscopic island forms; thus the
probability distribution is concentrated around the macroscopic value, which of
course scales as $L^2$.

\begin{figure}[htb]
\begin{center}
\includegraphics[width=12cm,angle=0]{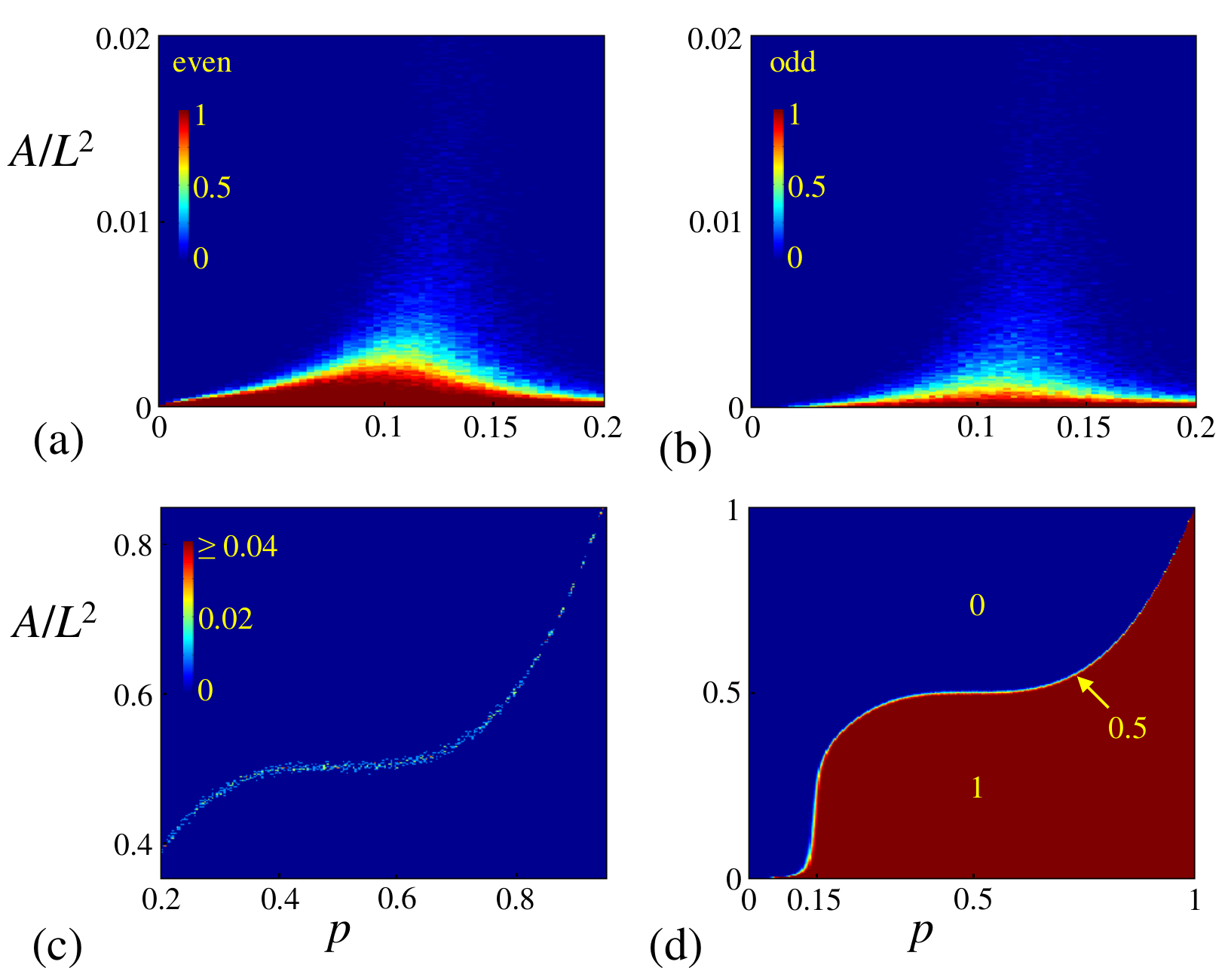}
\caption{ \label{Fig:area_distributions} %
The distribution of the normalized area $A/L^2$ of the N islands as a function of the
probability of flipping a hopping term $p$ for a $100 \times 100$ lattice. (a,b) The probability of having an island with area $A/L^2$ for small $p$. The probability is approximately exponentially distributed, in both even and odd values of $A$, regardless of the system size. This means that for small $p$ in every realization there are many microscopic islands. (c) For large $p$ the probability is concentrated around a mean macroscopic value, i.e. there is a single macro-island. (d) The probability of having an island with area larger than $A/L^2$ as a function of $p$. The curve of probability 0.5 indicates the characteristic largest island. The vertical slope at $\pc \approx 0.15$ indicates a percolation phase transition. The distribution was built upon 100 realizations. }
\end{center}
\end{figure}

This expectation is borne out by our numerical analysis. Figure \ref{Fig:area_distributions} depicts the distribution of the islands area $A$, normalized to the system's size $L^2$,
as a function of $p$ for an $L=100$ lattice. For large $p$ the distribution
appears to be concentrated around a mean value, which was numerically verified to
scale as $L^2$. In contrast, for small $p$ the distribution is approximately
exponential in a manner that was found to be independent of $L$. The transition takes
place at $\pc \approx 0.15$. The width of the transition between the microscopic and
macroscopic distributions, which takes place here in $0.10 < p < 0.15$, gets sharper
towards $\pc$ while increasing $L$.

Since flipping a hopping term flips the fluxes of two
triangles, islands with even values of $A$ are more common than islands with odd values of $A$.
Figure \ref{Fig:area_distributions} shows that in spite of the different
distribution of the even and odd values of $A$, both are exponentially distributed below $\pc$.

Figure \ref{Fig:area_distributions}(d) depicts, for every $p$ and normalized area $A/L^2$, the probability that there are islands larger in area than $A/L^2$. The curve for which this probability is $0.5$ is the typical size of the largest island for a particular disorder realization. We denote this curve by  $A_\mathrm{max}(p)$. If we measure the flux in the system with reference to the flux of the clean system, then since a triangle with N flux is formed by flipping either one or all three hopping terms forming the triangle, the mean flux threading the lattice is $\la \phi \ra = 2L^2[3p(1-p)^2 + p^3]$. Above $\pc$ the macroscopic island, if it exists, is supposed to capture almost all the N flux. Indeed, $A_\mathrm{max}(p > \pc)$ is excellently approximated by $\la \phi \ra$. Thus, above $\pc$ almost all the triangles with flipped flux are connected.

\begin{figure}[htb]
\begin{center}
\includegraphics[width=12cm,angle=0]{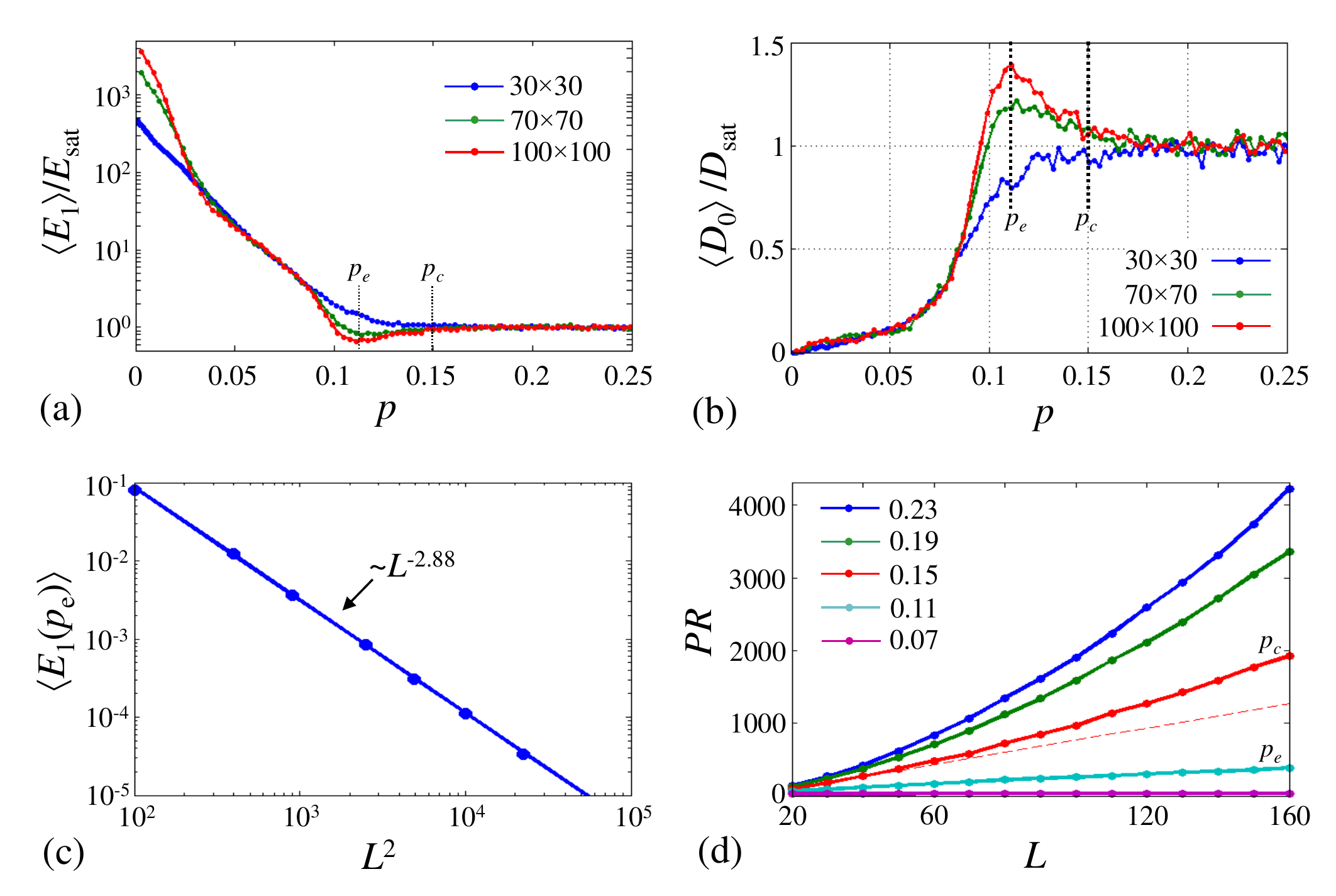}
\caption{ \label{Fig:D_and_Emin} %
(a) The normalized mean minimal energy $\la E_1 \ra / E_\mathrm{sat}$ as a function of  the probability $p$, where $E_\mathrm{sat} = \la E_1(p \rightarrow 0.5)\ra$. The percolation phase transition is manifested as the saturation to constant values
for $p > \pc$, but surprisingly the lowest energies appear around $\pe < \pc$, which
means that they belong to localized states. (b) The mean zero-energy DOS $\la D_0
\ra = \la D(0 \le E < E_\mathrm{sat}) \ra$ as a function of $p$, normalized by its value at $p=0.5$,  $D_\mathrm{sat}$, of the same
lattices. The saturation for $p > \pc$ is expected, but the DOS gets its maximal
values at $\pe < \pc$, and decays towards zero only in a polynomial manner. (c) The mean minimal energy at $\pe$ as a function of the system size is very well fitted to $L^{-2.88}$. (d) The disorder-averaged participation ratio $PR$ of the minimal energy state as a function of the system size $L$, for various values of $p$. For $p > \pc$ the state is extended, since $PR$ scales faster than $L$ (denoted by dashed line). For $p \ll \pc$ the states are localized, since $PR$ is independent of $L$. At $p \approx \pe$, $PR \sim L$, which implies a 1D string-like state. }
\end{center}
\end{figure}

Figure \ref{Fig:D_and_Emin} depicts two low-energy disorder-averaged quantities for various lattice sizes $L$. The first is $\la E_1(p) \ra$, the energy of the lowest excitation. The second is the zero-energy DOS $\la D_0 \ra$, which we define as the number of states within the energy window $0 \le E < E_\mathrm{sat}$, divided by $E_\mathrm{sat}$, where $E_\mathrm{sat} = \la E_1(p \rightarrow 0.5)\ra$ is the disorder averaged energy of the lowest excitation for the $p=0.5$ case, extracted from figure \ref{Fig:EnL}. In order to compare systems with different sizes, we normalize $\la E_1(p) \ra$ by $E_\mathrm{sat}(L)$, and $\la D_0 \ra$ by $D_\mathrm{sat} = \la D_0(p \rightarrow 0.5)\ra$.

For small $p$ we find $\la E_1\ra$ to approach the value of the energy gap, as it should. We find $\la D_0\ra$ to be polynomial in $p$ for small $p$, in accordance with our estimate for the probability of finding isolated regions of zero-energy states. Above the percolation phase transition the mean minimal energy $\la E_1 \ra$ and the zero-energy DOS $\la D_0 \ra$ are expected to saturate to a constant value as a
function of $p$.  We do indeed observe a saturation of $\la E_1\ra$ and $\la D_0\ra$ above $p_c$.

Surprisingly, figure \ref{Fig:D_and_Emin}(a) shows that the approach to the saturated value is not monotonic. The minimal $\la E_1 \ra$
appears at $\pe \approx 0.11 < \pc$, and we find $\la E_1(\pe) \ra \propto L^{-2.88}$, as shown in figure \ref{Fig:D_and_Emin}(c).
Moreover, the maximum $\la D_0 \ra$ also appear at $\pe$. The divergence of the DOS at zero energy seems therefore to be strongest before the percolation threshold.

If the percolation phase transition captures the essence of the low-energy
physics, we naively expect the low-energy states to be extended above $\pc$, while
those below $\pc$, including $\pe$, to be localized. For any eigenstate of the
Hamiltonian $\G_E = \sum_i a_i \g_i$, the localization properties of the state can be
characterized by the participation ratio (PR), defined by $PR = ( \sum_i |a_i|^4
)^{-1}$. The PR of a perfect metal scales linearly with the system size $L^2$, while the PR of an
insulator is independent of $L$. Figure \ref{Fig:D_and_Emin}(d) depicts
the disorder-averaged PR of the lowest excitation as a function of the system size,
for various values of $p$. As expected, above $\pc$ the PR scales faster than $L$, which indicates an extended state, while
well below $\pc$ the PR is independent on $L$. However, at $\pe$ the PR scales as
$L$. This scaling seems to imply that there is a range of $p$ in which the low-energy states are string-like in the sense that they are macroscopic only along one dimension. Recalling the observed diverging DOS of the edge states along the narrow strips, the divergence of $\la
D_0 \ra$ at $\pe$ is understood, even if not the precise exponent.


\section{Summary}
\label{Sec:Summary}

In the previous sections we have seen that the zero-energy peak of the DOS for
random signs in the nearest neighbor hopping terms may have different sources in two
limits of the probability $p$ for flipping the sign of individual hopping terms. For
small $p$ the peak is the consequence of the existence of localized zero-energy states and
the tunneling between them and other localized states. Therefore it appears on a
background of the zero DOS of the energy gap, and it is proportional to $p^3$. In
the opposite limit of $\pc \le p \le 0.5$, where $\pc \approx 0.15$, macroscopic
islands of the time-reversed phase cross the lattice, and the gap is closed due to
the low-energy modes along the interfaces of these islands with the original phase.
Since the dispersion of some of these modes is nonlinear, they are expected to
create zero-energy peak in the DOS. Our numerical results indicate a weak divergence
of the DOS at zero energy for $p=0.5$, and an even stronger divergence somewhat
below the percolation transition. In this region the islands seem to be of
the form of strings, and the mentioned nonlinear dispersion is likely to be the
source the divergence.

It was mentioned above that the Hamiltonian of our system is purely
imaginary and anti-symmetric both to charge conjugation and time-reversal. According
to the standard classification of disordered Hamiltonians \cite{AltlandZirnbauer} it
belongs to class D \cite{Mirlin_review1, Mirlin_review2}. This class is known to have three distinct phases: thermal
metal, thermal insulator and thermal quantum Hall (TQH) insulator
\cite{SenthilFisher,Bocquet}. In the metal phase, the DOS at zero energy diverges
logarithmically with the system size \cite{SenthilFisher}. Several numerical
studies of the properties of this class have been carried out using the Cho-Fisher network
model \cite{Chalker,Mildenberger1,Mildenberger2}, which exhibits all of the phase diagram.
It was found that in the TQH phase there is a region below the insulator-metal
transition where the DOS at zero energy diverges, but with a nonuniversal exponent,
due to the effects of rare configurations of disorder (Griffiths phase)
\cite{Mildenberger2}. Recently a model of MFs on a square lattice created in a $p$-wave
superconductor by a random potential has been proposed \cite{WimmerBeenakker}, and the
divergence in the metal phase has also been observed numerically there.

Our model is a new realization to the TQH--thermal metal transition, which
may be physically realizable. We observe the divergence of the DOS both in the metal
regime and in the analogue to the Griffiths regime, and the microscopic
understanding of our system provides insights into the physics of these phenomena.

{\it Note added:} After completing the preparation of this manuscript, we became
aware of a study of a similar system carried out by C Lohmann, AWW Ludwig and S
Trebst \cite{LohmannLudwig}.


\ack %
YEK thanks M Wimmer for providing his MATLAB implementation for the Pfaffian
\cite{WimmerPfaffian}. We thank AD Mirlin and CWJ Beenakker for pointing our attention to previous works. We also thank the US-Israel Binational Science Foundation, the Minerva foundation and Microsoft's station Q for financial support.

\appendix
\section{Minimal zero mode}
\label{App:Mininal}

In this appendix we prove that flipping three hopping terms with a common vertex in
a periodic clean lattice results in two localized states with energy which is either
zero or exponentially small with the system size. The proof also implies that there
is no way to create zero modes with flipping less than three signs.

The hopping signs $s_{ij}$ in the Hamiltonian (\ref{Eq:tightbinding}) define a
matrix $S$, which is skew-symmetric $S^T = -S$, and therefore has a well defined
Pfaffian. The recursive definition of the Pfaffian of an $N \times N$ matrix is
\begin{equation} \label{Eq:Pfaffian}
\Pf(A) = \sum_{i=1}^{N} (-1)^i a_{1i} \Pf(\hat{A}_{1i}), \\
\end{equation}
\begin{equation}
\Pf \left( \begin{array}{cc} 0 & a \\ -a & 0 \end{array} \right) = a \qquad \Pf(0) =
0, \nonumber
\end{equation}
where $a_{ij}$ is the element $i,j$ of $A$, and $\hat{A}_{1i}$ is the matrix $A$
without the $1^\mathrm{st}$ and $i^\mathrm{th}$ rows and column. A useful property
of the Pfaffian \cite{Nakahara} is that for any matrix $B$
\begin{equation} \label{Eq:DetPf}
\Pf(BAB^T) = \det(B) \cdot \Pf(A).
\end{equation}
The gauge transformation $\g_i \rightarrow -\g_i$ flips the signs of the $i^\mathrm{th}$ row and column of $S$. This transformation can be done by a diagonal matrix $B$ with the elements $b_{jj} = (-1)^{\delta_{ji}}$. And since $\det(B) = -1$, the gauge transformation flips the sign of $\Pf(S)$. Moreover, swapping labels of two sites, also flips the sign of $\Pf(S)$, since it can be
performed by $B = I_{(N-2)\times(N-2)} \times \sigma_x$, where $\sigma_x$ is the
Pauli matrix which acts in the space of the two swapped sites.

\begin{figure}[htb]
\begin{center}
\includegraphics[width=16cm,angle=0]{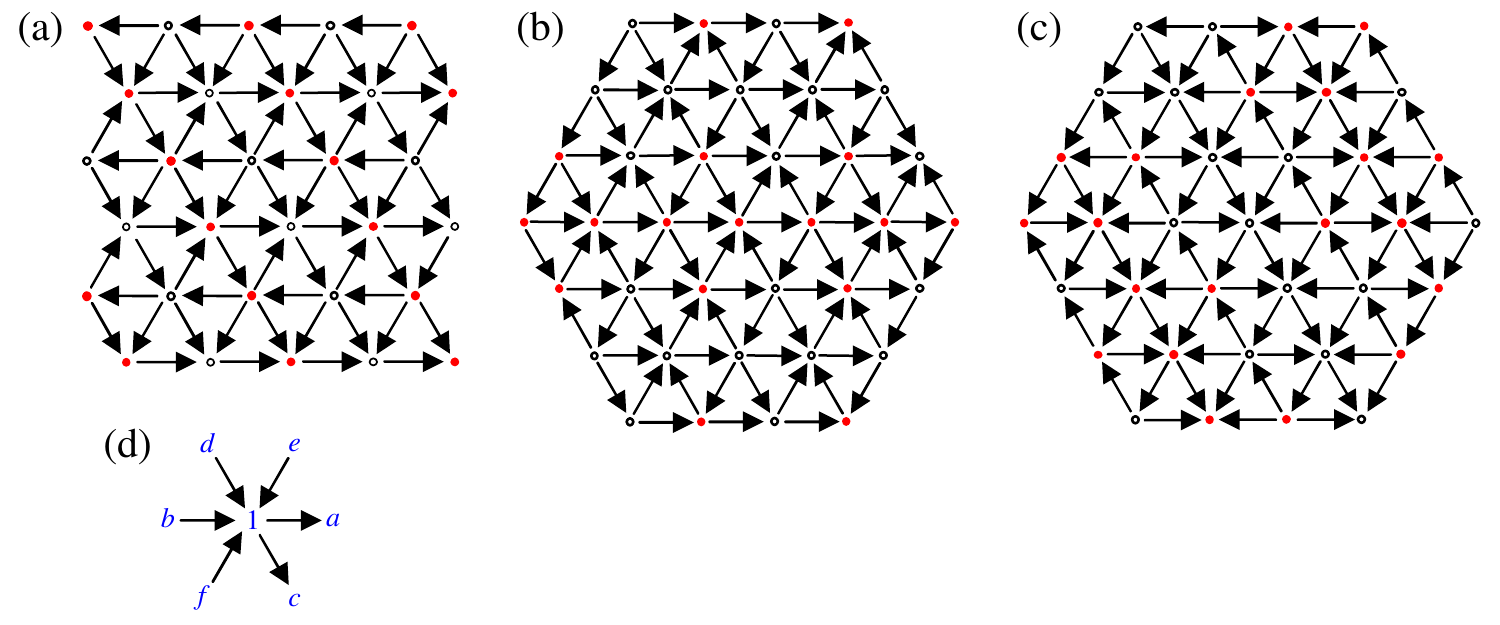}
\caption{ \label{Fig:lattice_gauges} %
(a) The gauges transformation that swaps between the two sublattices. A red cycle
means $\gamma_i \rightarrow -\g_i$; thus all the arrows connected to the site $i$
are flipped. (b,c) Gauge transformation which produce $60^\circ$ (b) clockwise and (c)
anticlockwise rotations. (d) An arbitrary site is denoted by 1 and its six neighbors
by $a$--$f$. }
\end{center}
\end{figure}

Systematic gauge transformations reveal equivalence relations between apparently
different lattices. The simplest example is the gauge transformation that swaps
between the two sublattices, which is depicted in figure
\ref{Fig:lattice_gauges}(a). In a periodic lattice the number of rows is even
(which means that the two sublattices are indeed equivalent), an even number of
sites are gauged, and the Pfaffian is unchanged. Rotations by $60^\circ$ clockwise
and counterclockwise can be also produced by gauge transformations, as depicted in
figures \ref{Fig:lattice_gauges}(b) and \ref{Fig:lattice_gauges}(c). In principle, such rotations may multiply the Pfaffian by a factor of $-1$. However, three $60^\circ$ rotations result in swapped
sublattices, which does not change the Pfaffian, and hence one rotation must leave the Pfaffian unchanged as well. Note that
in finite lattices the $60^\circ$ rotations may have $O(\e^{-L})$ corrections, due
to change of boundary conditions.

All these gauge-dependent properties of the Pfaffian show that it is an unphysical
quantity. However, since $\Pf(A)^2 = \det(A)$, the Hamiltonian $H$ has zero energy
states iff $\Pf(S) = 0$. Therefore the way to prove that flipping three neighboring
arrows gives zero-energy states, would be to prove that the Pfaffian vanishes for such a configuration.

Let us denote an arbitrary lattice site by 1, and its six neighbors by $a,b,..,f$,
as depicted in figure \ref{Fig:lattice_gauges}(d). According to definition
(\ref{Eq:Pfaffian})
\begin{equation} \label{Eq:Pf_S}
\Pf(S) = \sum_{i=a}^{f} (-1)^i s_{1i} \Pf(\hat{S}_{1i}).
\end{equation}
The physical meaning of $\Pf(\hat{S}_{1i})$ is omitting the sites 1 and $i$. We look
for relations between lattices with vacancies of two neighboring sites. These relations will allow us to express five of the terms $\Pf(\hat{S}_{1i})$ in (\ref{Eq:Pf_S}) in terms of the sixth, say $\Pf(\hat{S}_{1a})$.

Due to the horizontal symmetry of our gauge, it is apparent that
$|\Pf(\hat{S}_{1b})| = |\Pf(\hat{S}_{1a})|$, but since the labeling of the sites is
different, they may differ in signs. If, however, we `push' the $a^\mathrm{th}$ row
and column of $\hat{S}_{1b}$ $a-b-1$ rows and columns upwards, we get exactly
$\hat{S}_{1,a}$. This process involves $a-b-1$ labels swapping; thus
$\Pf(\hat{S}_{1b}) = (-1)^{a-b-1} \Pf(\hat{S}_{1a})$. $\hat{S}_{1c}$ and
$\hat{S}_{1d}$ are related by swapping the two sublattices; thus
$|\Pf(\hat{S}_{1d})| = |\Pf(\hat{S}_{1c})|$. By `pushing' the row and columns, we get
again $\Pf(\hat{S}_{1d}) = (-1)^{c-d-1} \Pf(\hat{S}_{1c})$. In a similar manner we
have $\Pf(\hat{S}_{1f}) = (-1)^{e-f} \Pf(\hat{S}_{1e})$. Furthermore, $\hat{S}_{1c}$
($\hat{S}_{1e}$) is related to $\hat{S}_{1a}$ by the anticlockwise (clockwise)
rotation, hence $|\Pf(\hat{S}_{1e})| \approx |\Pf(\hat{S}_{1c})| \approx
|\Pf(\hat{S}_{1a})|$. By counting the number of sites been gauges, we get
$\Pf(\hat{S}_{1c}) = (-1)^{a-c} \Pf(\hat{S}_{1a}) + O(\e^{-L})$ and
$\Pf(\hat{S}_{1e}) = (-1)^{a-e-1} \Pf(\hat{S}_{1a}) + O(\e^{-L})$. Substituting all
these relations in equation (\ref{Eq:Pf_S}) yields
\begin{equation} \label{Eq:6Pf_ia}
\Pf(S) = (-1)^a ( s_{1a} - s_{1b} + s_{1c} - s_{1d} - s_{1e} - s_{1f} ) \Pf(\hat{S}_{1a}) + O(\e^{-L}).
\end{equation}

In the clean lattice $s_{1a} = -s_{1b} = s_{1c} = -s_{1d} = -s_{1e} = -s_{1f}$, as can be seen in figure \ref{Fig:lattice_gauges}(d).
Thus the terms sum-up, and $|\Pf(S)| = 6 |\Pf(\hat{S}_{1a})|$. On the other hand, flipping three of the $s_{1i}$'s, gives
$\Pf(S) = 0$, i.e. zero-energy modes. Due to the locality of this manipulation, the
zero modes are localized around site 1. On the other hand, flipping only one or two
terms will not result in zero modes.


\end{document}